# New Experimental Method for Investigating AC-losses in Concentric HTS Power Cables

Steffen Elschner, Eduard Demencik, Bruno Douine, Francesco Grilli, Andrej Kudymow, Mark Stemmle, Severin Strauss, Victor Zermeno, Wilfried Goldacker

*Abstract*—The optimization of a HTS cable design with respect to AC-losses is of crucial importance for the economic viability of the respective concept. However the experimental determination of AC-losses is not straightforward since for short cable samples the distribution of current among the superconducting tapes is mainly determined by the contact resistances of the individual tapes. The resulting inhomogeneous current distribution definitely falsifies the results. To solve this experimental problem we present a new experimental technique. The setup is a 2m-long three phase concentric cable model for which, within each phase, the superconducting tapes (up to 30) are connected in series. The Cu-braid backwards conductors were assembled in a rotational symmetric cage type arrangement, such that their self fields at the cable cancel. If experimental peculiarities of this setup, as the strong inductive coupling between the phases and the suitable positioning of the voltage contact leads, are correctly taken into account, the currents can be controlled independently and the electrical properties of the cable can be measured unambiguously. In this paper preliminary results are presented. The work is part of the German government funded cable project AMPACITY (1 km / 20 kV/ 2 kA)

## I. INTRODUCTION

TRANSMISSION and distribution cables are today one of the most mature applications of HTS in electrical grids [1]. Recently a German government funded cable project, Ampacity [2], was accomplished with the installation of a 1 km, 10 kV, 2300 A, 3-phase concentric cable in the grid of RWE in Essen, Germany. Since April 2014 this cable has been operated successfully.

The cable is based on BSCCO 2223 HTS-conductors. The inner phase is equipped with 22 tapes in one layer wound on a former (outer diameter 36 mm) with a pitch angle of 16°.

The second and third phase have, due to the larger outer diameters, 26 and 30 tapes, with a similar pitch angle and a distance between the concentric phases of about 5 mm.

Since AC-losses are the essential part of the total thermal losses of a transmission cable, at least if it is operated near its critical current, their correct prediction is an important premise for the design of the cable. The corresponding values are decisive for the size of the needed cooling device, and thus crucial for the economic viability of the entire project. Since in general the transport losses of superconducting tapes are strongly dependent on current, $P \sim I^n$, $n=3…4$, a reduction is always possible by overdesigning the power carrying capacity. However, this would also increase costs due to larger cable diameters and an increased need of superconducting tapes.

For the installed Ampacity cable the AC-losses are not completely known up to present. A calorimetric determination on the 25 m-prototype-cable [3] leads to a high uncertainty, since the contribution of current leads still dominates at this length. For the now installed 1km-cable the needed steady state conditions are rarely realized within the real life grid.

An electric determination in both cases is prohibited, since the positioning of voltage leads inside the cryostat are not desired in the HV-environment. Moreover, the electric evaluation of AC-losses is particularly difficult in a three phase coaxial cable [4] as Ampacity. The magnetic fields at the surfaces of the tapes have axial and azimuthal components caused by the three independent currents. These fields are not in phase with the currents in the respective phases, which renders difficult both, experiment and simulation.

In order to evaluate the losses of the Ampacity cable by experiment we developed a special lab-sized setup which eliminates the current lead problems. This setup also allows the determination of the DC critical currents in individual tapes in the self-field of the cable (section 3). In sections 4 and 5 first AC-loss measurements are presented on a single phase and a two-phase configuration, respectively; the experimental challenges for the determination of losses in a three-phase system are discussed. In sect. 6 the losses for the Ampacity cable in the three-phase-configuration are estimated.

This work was supported partly by the German government, Grant Nr. 03ET1055D (Ampacity) and partly by the Helmholtz-University Young Investigator Group Grant VH-NG-617. *Corresponding author:* S. Elschner

S. Elschner is with University of Applied Science Mannheim, Mannheim, Germany, and with KIT, Karlsruhe, Germany, Corresponding author's e-mail: s.elschner@hs-mannheim.de. A. Kudymow, W. Goldacker, E. Demencik, S. Strauss, V. Zermeno, F. Grilli are with KIT, Karlsruhe, Germany. M. Stemmle is with Nexans Deutschland GmbH, Hannover, Germany. B. Douine is with GREEN Laboratory, Lorraine University, Vandoeuvre-lès-Nancy, France.

## II. Experimental Setup

The superconducting tapes used in Ampacity and in our experiments were manufactured by Sumitomo [5], multi-filamentary BSCCO 2223 in a silver matrix. The width is 4.8 mm, and the thickness about 0.4 mm, including 100 µm of Cu-alloy-stabilization. The mean gap between tapes was about 0.5 mm, i.e. 1 mm between superconducting ellipses.

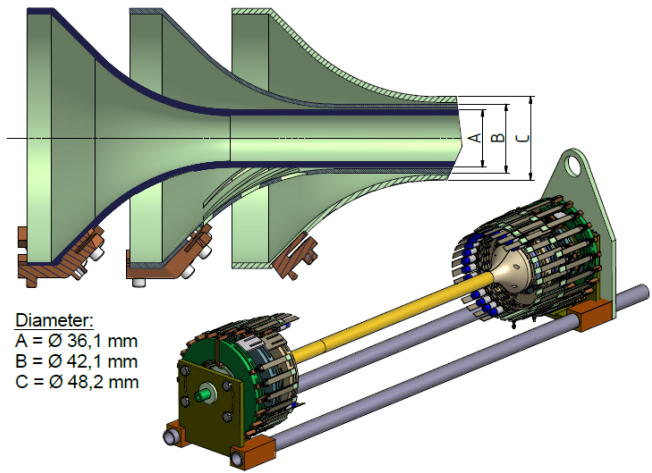

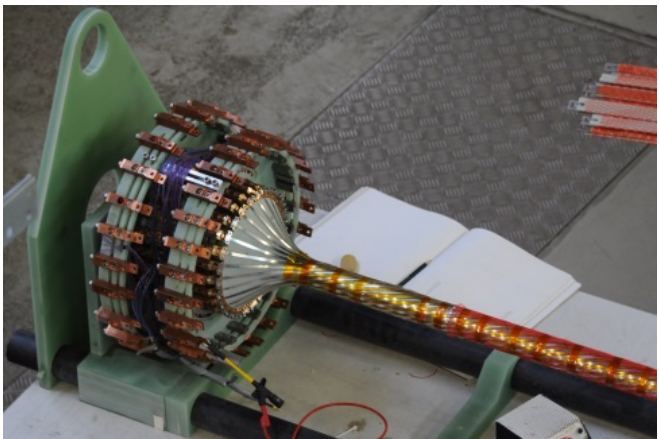

**Fig. 1:** Experimental setup for serial tape connection, schematic and photo, 2 phases installed, backwards conductors interrupted / not mounted.

For the experimental determination of losses in cables in principle the calorimetric method is preferable. For this method there is no doubt with respect to the interpretation of the results. However for short (~m) cable samples this method is not realistic, since then the losses are dominated by the contribution of current leads. Also electric methods are strongly hampered in short test specimen. The current distribution within the parallel tapes of each phase is then determined by the contact resistances of the individual tapes. This results in an inhomogeneous current distribution, even phase differences occur [6], and the loss determination becomes accidental. For long cables the impact of this effect vanishes, since the inductive contributions to the impedance and their coupling, equal for all tapes, dominate and equalize the currents among the tapes.

Our experimental concept is based on a serial connection of the superconducting tapes [7]. The tapes are positioned on the former exactly as in the Ampacity-cable but the tapes of each phase are connected in series with the current flowing in the same direction. The backwards conductors are arranged in a distance of about 25 cm in a cage type coaxial geometry. Due to the large number of backwards conductors, 22, 26, 30 for the three phases, this cage type arrangement behaves similarly to a homogeneous cylinder, i.e. is nearly field free in its inner volume. The superconducting tapes of the cable thus feel negligible field from the backwards conductors and their losses are only determined by the self field of the cable itself, as in the real device.

Indeed the residual magnetic field of the backwards conductors at the location of the tapes can be evaluated to less than 0.2 mT compared to self fields in the order of 40 mT. In order to reduce eddy losses in the copper backwards conductors we chose copper braid (20 x 5 mm²).

Our experimental set-up is configured for a 3-phase installation (Fig. 1), but up to now only measurements in a 2-phase configuration have been performed. Diameters (36 and 42 mm) and pitch lengths (400 mm) were chosen as in the Ampacity cable, the winding direction of the spirals was the same in both phases.

The whole structure (2.5 m x 0.6 m x 0.6 m) was then immersed in a large polysterene tub filled with liquid nitrogen. All experiments were performed at T=77 K.

The AC-losses were determined by phase sensitive measurement of current and voltage. The current in the serial connection, phase and amplitude, is measured with calibrated Rogowski coils. For the recording of the voltage we attached lead wires with a low melting solder on the silver-matrix of BSCCO-tapes. The distance between voltage taps was exactly two pitch lengths, i.e. 800 mm. The losses were then determined by two different methods which both brought about the same results, measuring with a commercial Lock-In - amplifier with the current as a reference or integrating the product $u(t) \times i(t)$ over one AC-cycle.

It is well established, that the resulting AC-losses strongly depend on the exact positioning of the voltage loops [8][9]. This has been shown by experiment [10] and numerical simulation [11] for elliptic superconducting cross sections. If the voltage leads are arranged along the center of the top plane of the tape, the losses are strongly under-evaluated, if they are arranged along the edge, slightly overestimated [11]. With increasing distance from the tape the results in both cases tend to approach the correct value. On the other hand the inductive signals increase with increasing loop size and can be up to 2-3 orders of magnitude larger than the loss signal, such that a compromise has to be found.

In our experiments we used three types of voltage loops (see Fig. 2).

a) *Helical loops*, the measuring leads are positioned parallel to the superconducting tapes in the center of their top plane. Including the isolation of the wire and the normal conducting component of the tapes the mean distance between wires and center of the tapes was 0.7 mm. According to [10][11], we expect an underestimation of losses.

b) *Axial loops*, the measuring wires are led parallel to the axis of the cable on top of the wires. The voltage wires are nearly everywhere far away from the tape, therefore we expect correct results. However the loop surfaces are large and increased scatter is expected.

c) *s-shaped loops*, the measuring wires follow the tape (as in the helical configuration), but they are led in the gaps

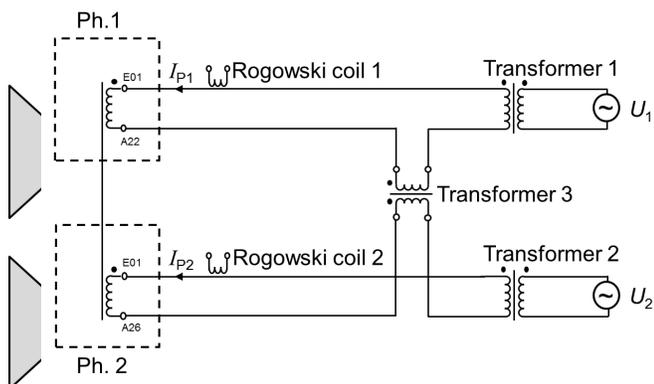

**Fig. 3:** electrical circuit of the experimental set-up (current generation)

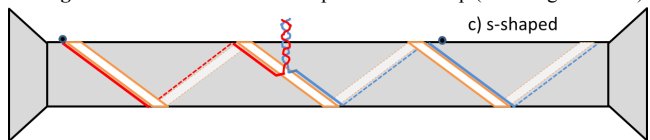

**Fig. 2:** Helical (a), axial (b) and *s*-shaped (c) loops used for AC loss measurements. For simplicity, only one tape is shown in the schematic diagrams.

between tapes [12]. In order to compensate external fields at least partly, the wires cross the tape in the middle between both voltage taps. We expect a slight overestimation of the losses (edge type). For design considerations such a slight overestimation is better acceptable.

The currents in both phases were induced with transformers (transformers 1 and 2 in Fig. 3). The sinusoidal output of a frequency generator with frequencies between 8 and 80 Hz was given on the primary side of a transformer with 50:1 turns to generate the desired large current in the cable.

In the operation mode with two phases an additional experimental problem emerged. Both phases with their respective backward conductors constitute a strongly coupled toroidal coil system. The mutual inductive coupling strongly influences the phases in both current systems.

For compensation of this effect an additional transformer (transformer 3 in Fig. 3) was introduced, with the same ratio of turns as the two phases, i.e. 22 and 26). However, even with this compensation the relative phases of both currents could not be adjusted with the desired precision, therefore the transformer 3 was equipped with an additional active winding which makes a fine tuning of the relative phase possible.

With this method the recorded Rogowski signals differed by less than 0.2°. In our experiments we investigated the relative phases $\Delta\varphi = 0°$ and $\Delta\varphi = 180°$, but it was also possible to adjust a phase relation $\Delta\varphi = 120°$. Measurements in the situation $\Delta\varphi = 120°$ up to now were not possible, since the compensation of the large inductive signals in this situation is not straightforward and remains a challenge.

## III. DC-Experiments

In a first experiment we determined the DC critical current of the tapes alone and in the cable with the usual 1 µV/cm criterion. The 22 tapes of the inner phase were all positioned on the former (R=18.1 mm) in the same geometry as in the Ampacity cable. In a first step the critical current of the tapes alone was evaluated, without current in the other tapes. We obtained an average (9 tapes) of $I_c$ = 168 A (red asterisks) in Fig. 4). In a second step the tapes were connected in series. Again the critical current was measured for each tape, we obtained a mean value of $I_c$ = 180 A (blue crosses), an increase of about 7 %.

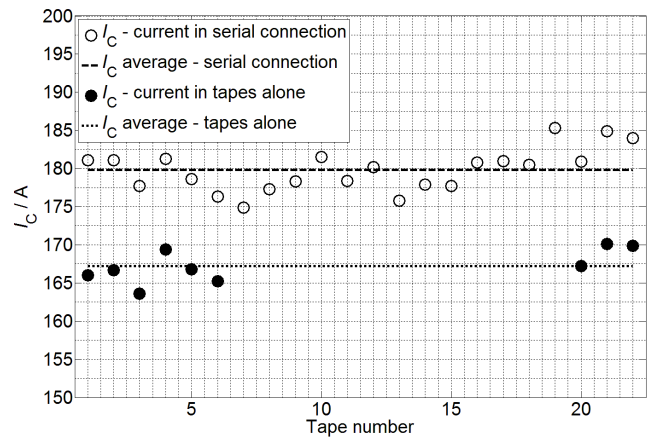

**Fig. 4:** DC critical current in single tapes and in the cable.

At a first glance this result seems astonishing since the magnetic self-field at the surface of the tapes alone is about $B \cong \mu_0 \cdot I / 2b \cong 26$ mT, somewhat smaller compared to the cable $B \cong \mu_0 \cdot 22 \cdot I / 2\pi R \cong 44$ mT. However this effect is at least qualitatively explained by the orientation of the magnetic fields. For the tape alone the self-field has, at the edges, strong components perpendicular to the tape whereas in the cable the self-field is nearly azimuthal with mainly parallel components, see numerical simulations in Refs. [13][14].

This experiment has then been extended to the two-phase configuration. First the critical currents of the 26 tapes of phase 2 were measured in the tape alone and serial configuration. Again we obtain an increase from $I_c$ = 168 A to 181 A with excellent reproducibility. If in addition a current flows through the 22 tapes in the inner phase 1 the critical current of the tapes in phase 2 depends of the direction of that current. At a current of $I$ = 170 A in phase 1, i.e. near $I_c$, the critical current in phase 2 is reduced to $I_c$ = 175 A if the currents have the same direction and is increased to $I_c$ = 186 A if the currents have opposite direction, always a mean values of the 26 tapes, again with excellent (± 2 A) reproducibility. Both results are qualitatively understood as an effect of local field.

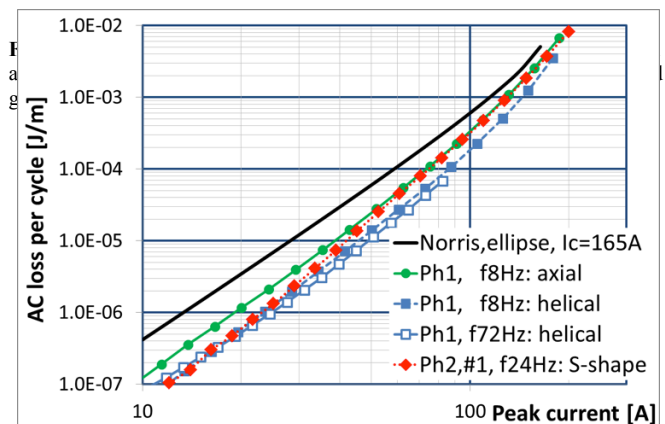

## IV. AC-LOSSES SINGLE PHASE

The AC-losses of a single (the inner) phase alone were determined with different experimental methods. Most of the results were obtained with the lock-in-technique but it was always checked that numerical integration over one cycle gave the same results. In Fig. 5 the experimental results are gathered. Plotted are always the losses per cycle and unit length for one tape versus the amplitude of the current. To guide the eye, the losses of a Norris ellipse [15] is also depicted.

To verify that indeed only AC-losses were measured we checked that the losses per cycle were independent of the measuring frequency. For this check we chose the helical voltage geometry. Since in this geometry the AC-losses are underestimated (see section 2) it is expected that parasite effects not proportional to the frequency, e.g. eddy currents in the Cu-stabilization, are particularly well visible. However a comparison of results at 8.1 Hz (blue full square symbols) and at 72 Hz (blue open square symbols) did not show any significant difference. The measurements at 72 Hz were only conducted up to a current of 80 $A_p$ because of the limitations of the current source. The depicted results are the mean value of the 22 tapes, with a scatter in the order of ±10%. If we assume that the average distance of the voltage leads from the center of the sample is about y = 0.7 mm and the long half axis of the superconducting ellipse about a = 2mm, we obtain a ratio y/a = 0.35, and from the simulations of [11] we expect an underestimation of the correct losses by about 30%.

This underestimation of AC-losses becomes obvious from comparison with the results obtained with axial loop geometry. In this geometry we expect exact results. As a frequency we chose again 8.1 Hz (green circles), plotted is again a mean value of the 22 tapes. However, due to the large measuring loop, the deviation of the results from the mean value is quite large ± 40%. Nevertheless the mean value indeed shows losses increased by about 30 % with respect to the helical measuring geometry.

On one of the 22 tapes (tape #1) an additional s-shaped voltage loop was connected and measured. The results (red diamond symbols line) coincide with the axial geometry.

From the AC-loss experiments on a single phase it can be concluded, that the results depend on voltage geometry as expected, and confirm that the losses in the tapes of a cable are smaller than those in a single ellipse in Norris-approximation [16]. Again this is explained by reduced radial field components in the cable [17].

## V. AC-LOSSES, TWO PHASES

In a further stage of the experiments a second phase was mounted (radius of the former: 21.1 mm, 26 tapes). The voltage was measured on one arbitrarily chosen tape of the outer phase, the voltage geometry was s-shaped, i.e. the losses are slightly overestimated. In this 2-phase configuration two experiments were performed.

In a first experiment the currents in both phases were in phase, i.e. Δφ = 0. This is the situation as is present in a multilayer single phase cable. The amplitudes were such that the total amplitude in both phases was equal, i.e. the currents in the serial connections have the ratio of the number of tapes (i.e. 22/26).

From the results in Fig. 6 it becomes obvious that the losses in the outer phase are strongly increased by a factor of more than two compared to the situation of zero current in phase 1. This is plausible, since both fields sum up, and the oscillating fields at the surfaces of the tape have increased amplitude [18].

In a second experiment the relative phase of both currents was chosen Δφ = 180°, i.e. the current in both phases was antiparallel. This situation corresponds to a bipolar coaxial cable, or to the external phase of a three-phase coaxial cable as Ampacity. Compared to the situation without current in the inner phase the losses are slightly decreased. Also this result is qualitatively expected because of partial field compensation.

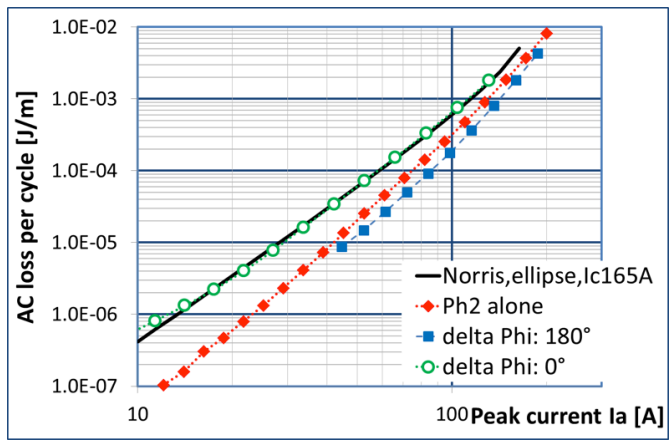

**Fig. 6:** AC-losses im Phase 2 in the Δφ =0°- configuration (green circles), in the Δφ =180°- configuration (blue squares) and without current in phase 1 (red diamonds), tape 1#, S-shaped voltage loop.

## VI. Losses of Ampacity cable

A precise prediction of the total losses of Ampacity cable is not yet possible with the presented results. However an approximate evaluation can be attempted. In Ampacity the axial field effects probably can be neglected. This is e.g. seen in simulation work [19], where with a ratio between pitch length and diameter of 10, as in Ampacity, the effect of pitch to the losses is less than 10 %.

The losses of the inner phase are then well represented by the situation of a single phase, as measured in section 4. At the nominal current of 2300 $A_{eff}$ (i.e. 147 $A_p$ per tape) we obtain from Fig. 5 $Q$ = 2 mJ/cycle/m. At 50 Hz and with 22 tapes this gives $P$ = 2.2 W/m.

In the outer phase (30 tapes) the current per tape is only 109 $A_p$. Since the field at the outer phase corresponds exactly to a 2-phase cable with a configuration of Δφ = 180° we read from Fig. 6 $Q$ = 0.4 mJ/ cycle/m, which gives P = 0.6 W/m for the outer phase.

For the losses in the middle phase (26 tapes, i.e. 125 $A_p$) measurements at a phase relation of Δφ = 120° are needed. Although this phase relation can be adjusted with good accuracy for the current, such measurements have not been possible up to now (section 2). However we can argue, that the losses are somewhere in between the losses at Δφ = 0 and Δφ = 180°. An upper limit for the losses in the middle phase thus is again $Q$ = 2 mJ/cycle/m, i.e. $P$ = 2.6 W/m.

As a result we predict losses of $P$ = 5.4 W/m for the entire 3-phase cable at nominal current and $T$ = 77 K. However, in the actual long cable the temperature depends on the position (68 K – 77 K) such that the mean losses will be sensitively smaller.

## VII. Summary

In this paper, we have presented a novel method for measuring AC-losses in a coaxial 3-phase cable, which avoids the well known problems with current distribution in short cable samples. AC-losses have been measured for the one- and two-phase configurations, the latter with variable phase shift between the phases. In the one-phase case, the losses are lower than the corresponding ones for single tapes, due to the partial cancelation of the perpendicular field component impinging on the tapes. In the two-phase case, the losses strongly depend on the relative phase, as expected from intuitive considerations on the interaction between the magnetic fields generated by the two superconducting layers. We plan to use this experimental set-up to investigate the AC-loss behavior of other cable configurations (e.g. two and three phases shifted by 120°) and composing superconducting tapes (e.g. HTS coated conductors).